\documentclass[preprint,aps,amsmath,amssymb,superscriptaddress]{revtex4-1}

\usepackage{graphicx}
\usepackage{dcolumn}
\usepackage{bm}
\usepackage{xcolor}

\begin{document}


\title{High-order sideband effect with two-level atoms in a shaken optical lattice}


\author{Mo-Juan Yin$^{\dag}$}
\affiliation{CAS Key Laboratory of Time and Frequency Primary Standards, National Time Service Center, Xi¡¯an 710600, China}
\author{Jie-Yun Yan$^{*,\dag}$}
\affiliation{School of Science, Beijing University of Posts and Telecommunications, Beijing 100876, China}
\author{Qin-Fang Xu$^{\dag}$}
\affiliation{CAS Key Laboratory of Time and Frequency Primary Standards, National Time Service Center, Xi¡¯an 710600, China}
\author{Yang Guo}
\affiliation{CAS Key Laboratory of Time and Frequency Primary Standards, National Time Service Center, Xi¡¯an 710600, China}
\author{Ben-Quan Lu}
\affiliation{CAS Key Laboratory of Time and Frequency Primary Standards, National Time Service Center, Xi¡¯an 710600, China}
\author{Jie Ren}
\affiliation{CAS Key Laboratory of Time and Frequency Primary Standards, National Time Service Center, Xi¡¯an 710600, China}
\author{Ye-Bing Wang}
\affiliation{CAS Key Laboratory of Time and Frequency Primary Standards, National Time Service Center, Xi¡¯an 710600, China}
\author{Lin-Xiang He}
\affiliation{State Key Laboratory of Magnetic Resonance and Atomic and Molecular Physics, Wuhan Institute of Physics and Mathematics, Chinese Academy of Sciences, Wuhan 430071, China}
\author{Min Xiao$^{*}$}
\affiliation{National Laboratory of Solid State Microstructures and School of Physics, Nanjing University, Nanjing 210093, China}
\affiliation{Department of Physics, University of Arkansas, Fayetteville, Arkansas 72701, USA}
\author{Hong Chang$^{*}$}
\affiliation{CAS Key Laboratory of Time and Frequency Primary Standards, National Time Service Center, Xi¡¯an 710600, China}

\begin{abstract}
The system of cold two-level atoms inside an optical lattice potential has been used to simulate various models encountered in fundamental and condensed-matter physics. When the optical lattice is periodically shaken, some interesting dynamical effects were observed. In this work, by utilizing the ultra-narrow linewidth of the dipole-forbidden transition in $^{87}$Sr atoms loaded in a shaken optical lattice, we reach a new operation regime where the shaking rate is much larger than linewidth of the two-level transition but much smaller than the frequency interval of the harmonic lattice potential, which was not realizable before. Under such unique conditions, an interesting quantum dynamical effect, the high-order sideband effect (HSE), has been experimentally observed. All the results can be well explained by a developed theoretical model with analytical solution. This HSE is governed by different mechanism as the similar phenomenon observed in semiconductors and has unique characteristics.
\end{abstract}



\maketitle

Two-level atomic systems are the fundamental building blocks in physics \cite{Allen1975}. Many interesting physics phenomena arouse when such simple physical systems interact with electromagnetic fields or are placed in various external environments. A system with ultracold two-level atoms loaded into a one-dimensional optical lattice potential formed by a standing-wave field can be used to model many novel problems encountered in condensed-matter physics.\cite{Eckardt2017, Greiner2008, Bloch2008, Goldman2016, Gadway2012, Kolkowitz2017} When the optical lattice is modulated (or shaken) with a certain rate by introducing a frequency difference or time-varying phase difference between the two laser fields forming the standing-wave potential as commonly used, more interesting dynamical processes could occur. \cite{Lignier2007, Lundblad2008, Tokuno2011, Struck2012, Parker2013, Ha2015, Clark2016} So far, all experimental works on two-level systems in shaken optical lattices have been conducted in a regime where the lattice shaking rate could modify the band structure of the lattice potential. Thus, the opposite regime, in which the lattice shaking rate is small enough to avoid disturbing the energy structure of the lattice trap potential but still much larger than the decay rate of the two-level system, should represent an interesting research subject. However, the above conditions are difficult to fulfil experimentally because the decay rates of most two-level atoms \cite{Gutterres2002, Sell2011} and solid-sate systems for optical dipole-allowed transitions are typically large. In such a case it is impossible to keep the lattice shaking rate to be larger than the decay rate of the two-level system but smaller than the energy gap (level difference) of the optical lattice trap potential (on the order of tens kHz).
	
In this Letter, we experimentally demonstrate a new system consisted of a shaken one-dimensional optical lattice loaded with ultracold $^{87}$Sr atoms, which can work in the regime of having the shaking rate being far lower than needed to change the energy structure but much larger than the atomic decay rate to permit certain interesting dynamical processes. By taking advantage of the ultralong lifetime of the selected dipole-forbidden transition in the ultracold $^{87}$Sr atom, a perfect two-level atomic system with a measured linewidth as low as $8.6\times2\pi$ Hz can be realized. This two-level transition rate is far more narrower than the best performances of semiconductor quantum dots and solid-sate systems \cite{Juska2013, Keil2017}, as well as atoms with optical dipole-allowed transitions. In this system, the optical lattice can be shaken at a rate on the order of hundreds of Hz, which is much larger than the linewidth of two-level systems ($<$10 Hz) but still small enough to avoid any effective coupling between the energy levels of the lattice potential. Such a lattice shaking rate can protect the perfect two-level system in the lattice and facilitate investigations of the novel dynamical behaviours of the two-level system in this new platform.
	
Using this platform with ideal two-level atoms inside the shaken periodic potential, we have experimentally observed an interesting quantum dynamical phenomenon, i.e., a high-order sideband effect of the two-level atomic transition. To observe a high-order sideband effect, the atomic transition must be modulated with a strong ac field. Ingeniously configuring the shaken optical lattice and the interacting lights, an equivalent ac field is introduced to act on the internal transition of the two-level atoms by the Doppler effect resulting from the external motion of the atoms in the moving lattice frame. Interestingly, the strength of the effective ac field is determined by the modulation amplitude of the laser frequency that forms the periodic potential in the experiment. The observed high-order sidebands accompanying the otherwise single peak of the resonant transition expand the output spectrum with multiple peaks having the same frequency interval of shaking rate, which are distributed symmetrically on both sides of the central resonance peak. When the modulation amplitude is changed, the total number and the relative intensities of the sidebands also change significantly. A theoretical model has been developed to describe this simple system with an exact analytical solution that can well explain all the experimental observations. Although similar high-order sideband generation (HSG) effect was theoretically proposed \cite{Liu2007, Yan2008} and observed experimentally in semiconductors \cite{Zaks2012, Zaks2013, Langer2016} recently, certain properties that are easily observed in our current atomic experiment cannot be observed with the accelerated excitons in semiconductors. Also, to reach the required ac field strength for observing such HSG effect in semiconductors, free electron laser was needed, which significantly limits its applicability in more general systems. By comparison, large ac field strength can be achieved simply by increasing the frequency modulation amplitude of our diode laser used to form the optical lattice. The attainable stability, tunability, and measurement precision in our atomic experiment are difficult to achieve in the solid-state counterparts because of the large decay rates and strong required acceleration field. The ultra-narrow linewidth of this perfect two-level atomic system and the easily-modulated periodic potential make the current system be an ideal platform to further investigate many novel effects predicted with this simple two-level atom inside a shaken optical lattice. For example, a quantum computation scheme was recently proposed in the shaken optical lattice system, \cite{Schneider2012} in which the generated high-order sidebands could be used as a multiplexer to significantly increase the computational capability.

\begin{figure}[ptb]
\centering
\includegraphics[width=8.5cm]{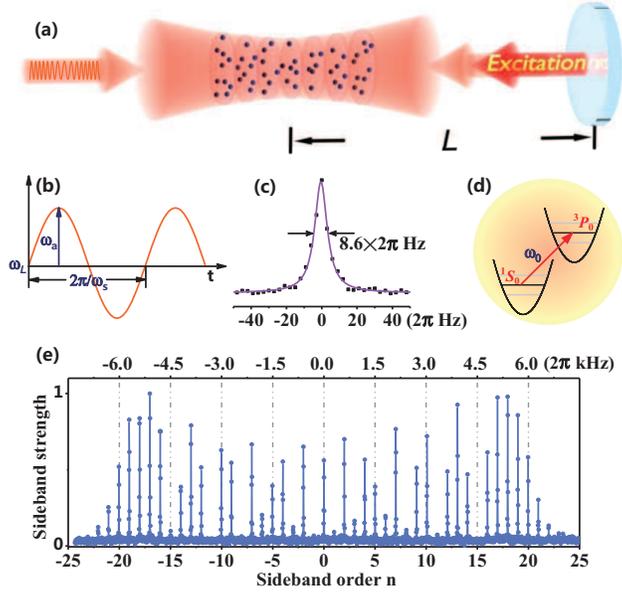}
\caption{HSE in the shaken optical lattice. (a) Schematic diagram of two-level atoms in the shaken optical lattice. Both the frequency-modulated lattice light centered at 813 nm and the excitation laser centered at 698 nm propagate along the same axis. The lattice light enters from left and is reflected by the mirror to form the standing-wave periodic potential. The excitation laser beam is injected from the right through the mirror. The length of the atomic sample is about 1 mm. The distance between the formed optical lattice and the mirror is labelled as. (b) An illustration of defined parameters for the frequency modulation of the lattice beam. (c) The measured linewidth of the transition $^1$$S_0$-$^3$$P_0$ in the unshaken optical lattice is about $8.6\times2\pi$ Hz, with the excitation power of about 150 nW. (d) The energy structure of cold atoms in the trapping optical lattice, which consists of a series of equidistant levels in the approximate harmonic potential. The arrow denotes the dipole forbidden transition $^1$$S_0$-$^3$$P_0$ in $^{87}$Sr. (e) The typical high-order sideband spectrum measured with two-level atoms in the shaken optical lattice under the parameters: $\omega_s$ is $300\times2\pi$ Hz, $\omega_a$ is about $2.7\times2\pi$ GHz, the power of lattice light is 300 mW, and the power of excitation is 3 $\mu$W. The intensities of sidebands are normalized to the maximal peak. The sideband orders are shown at the bottom with the corresponding frequencies in unit of $2\pi$ kHz given on the top axis.}
\label{figure1}
\end{figure}

To prepare the system, we load the ultracold fermionic $^{87}$Sr atoms into the one-dimensional optical lattice formed by an 813 nm light beam retro-reflecting itself from a fixed mirror (Fig. \ref{figure1}a). The dipole-forbidden transition $^1$$S_0$-$^3$$P_0$ of the $^{87}$Sr atom \cite{Takamoto2003, Campbell2017} is chosen as the closed two-level system, which is excited by an excitation light around its resonant wavelength of 698 nm. The response of the system to the near-resonance excitation is obtained by measuring the excitation fraction of the two-state system using the normalized detection method \cite{Bergquist1986}. The measured linewidth in the transition spectrum is down to $8.6\times2\pi$ Hz (Fig. \ref{figure1}c). The standing-wave trapping field confines the atoms in a periodic harmonic potential (Fig. \ref{figure1}d) with an energy level interval of $\hbar\omega_p$. The estimated gap frequency $\omega_p$ is $65\times2\pi$ kHz, which is obtained by employing the resolved sideband spectroscopy of the transition in the unshaken optical lattice (see Supplementary Information). Then, the optical lattice is shaken by frequency-modulating the standing-wave light field as follows(Fig. \ref{figure1}b):
\begin{equation}
  \omega(t)=\omega_L+\omega_a\sin\omega_s t.
\end{equation}
where $\omega_L$ is the frequency of the 813 nm laser, $\omega_a$ denotes the amplitude of the frequency modulation, and $\omega_s$ is the shaking rate. The potential function of the shaken optical lattice is determined approximately by (see S.I.)
\begin{equation}
\cos^2\left[\frac{\omega_L}{c}+\frac{\omega_aL}{c}\sin\omega_st\right],
\end{equation}
where $c$ is the vacuum speed of light. Due to the shaking, the optical lattice experiences a time-dependent position shift of $\delta x(t)=\frac{\omega_a}{\omega_L}L\sin\omega_st$ in the laboratory frame, which means that the atoms gain an additional velocity of $v(t)=\frac{\omega_a\omega_s}{\omega_L}L\cos\omega_st$ in the moving optical lattice frame. Since the shaking rate $\omega_s$ (in tens or hundreds Hz in our experiment) is far smaller than $\omega_p$, it will not bring effective coupling or modification to the different energy levels in the lattice potential. However, the shaken optical lattice has an important influence on the two-level atomic transition. To see it more clearly, we now consider the two-level transition in the frame of moving optical lattice, where the potential is static and the atoms have a velocity relative to the excitation laser beam as analyzed before. This velocity will induce a frequency shift for the transition caused by the Doppler effect because the excitation light propagates along the lattice shaking direction and the transition linewidth is very narrow. Labelling the original two-level transition energy as $\hbar\omega_0$, the modified Hamiltonian of the $^1$$S_0$-$^3$$P_0$ transition is expressed as follows:
\begin{equation}
  H(t)=\hbar\omega_0+F\cos\omega_st,\label{Hamiltonian}
\end{equation}
where $F=\alpha\hbar\omega_s$ with $\alpha$ representing a dimensionless quantity given by $\alpha=\frac{\omega_a\omega_0L}{\omega_Lc}$. Thus, the system is constructed for driving two-level atoms by a time-dependent external field via periodically shaking the optical lattice.

If the transition is excited with photons having frequency $\omega_e=\omega_0$, then the high-order sideband generation at frequency $\omega_d$ is governed by the response function $\chi(\omega_d,\omega_e)$, which predicts to have $\omega_d=\omega_0+n\omega_s$  (n is an integer). In this experiment, we excite the two-level transition with a varying photon frequency $\omega_e$, and then detect the fractional population of the excited state \cite{Bergquist1986} at the fixed frequency of $\omega_d=\omega_0$. It can be shown that $\chi(\omega_0,\omega_0+n\omega_s)=\chi(\omega_0+n\omega_s,\omega_0)$. We label the generated peak at $\omega_0+n\omega_s$ as the $n$-th sideband on the measured spectrum.

The high-order sideband effect can be observed in our system for the shaking rate $\omega_s$ ranging from $10\times2\pi$ to $500\times2\pi$ Hz with a modulation amplitude $\omega_a$ on the order of $0.1-10\times2\pi$ GHz. Figure \ref{figure1}(e) shows a typical high-order sideband spectrum measured with a shaking rate of $\omega_s=300\times2\pi$ Hz, where at least ¡À20 sidebands are clearly observed.

In theory, we can analytically solve Eq. (\ref{Hamiltonian}) to obtain the relative strengths of the sideband peaks as (see S.I.)
\begin{equation}
\chi(\omega_0,\omega_0+n\omega_s)=
2\pi\mu^2 i^{3n+1}\int_0^{+\infty}
e^{i(\frac{n\omega_s}{2}+i\gamma_2)\tau}
J_{n}\left(\frac{2F}{\hbar\omega_s}\sin\frac{\omega_s\tau}{2}\right)d\tau.
\label{sideband}
\end{equation}
where $J_n(\cdots)$ is the $n$-th order Bessel function of the first kind. It is easy to see that the distribution of relative strengths on the sidebands is independent of the shaking rate $\omega_s$. In another word, the sideband spectra have the same number and strength distribution of the sidebands for different shaking rates, although the sideband interval $\omega_s$ changes synchronously.

\begin{figure}[ptb]
\centering
\includegraphics[width=8.5cm]{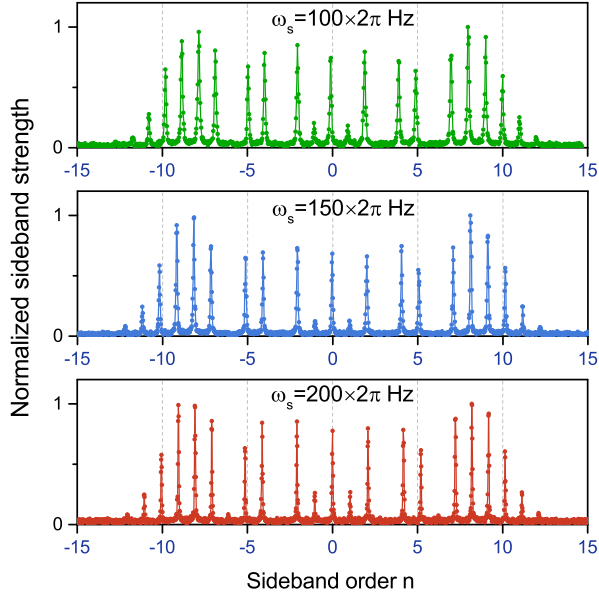}
\caption{The high-order sideband spectra for different shaking rates. The shaking rate $\omega_s$ sets at $100\times2\pi$, $150\times2\pi$, and $200\times2\pi$ Hz, respectively. The modulation amplitude $\omega_a$ is fixed at $1.3\times2\pi$ GHz. Power of the excitation light is about 3 $\mu$W.}
\label{figure2}
\end{figure}

Figure \ref{figure2} shows the high-order sideband spectra for different shaking rates (at $100\times2\pi$, $150\times2\pi$, and $200\times2\pi$ Hz from top to bottom). The experimental results clearly verify that the high-order sideband spectra are approximately equivalent when the value of $\omega_a$ is fixed and $\omega_s$ is changed within a certain range. The interval between neighboring sideband peaks actually changes with $\omega_s$. Since we have plotted the horizontal axis for each spectrum in terms of the sideband orders, each spectrum actually has a different frequency scale in Fig. \ref{figure2}.

\begin{figure}[ptb]
\centering
\includegraphics[width=8.5cm]{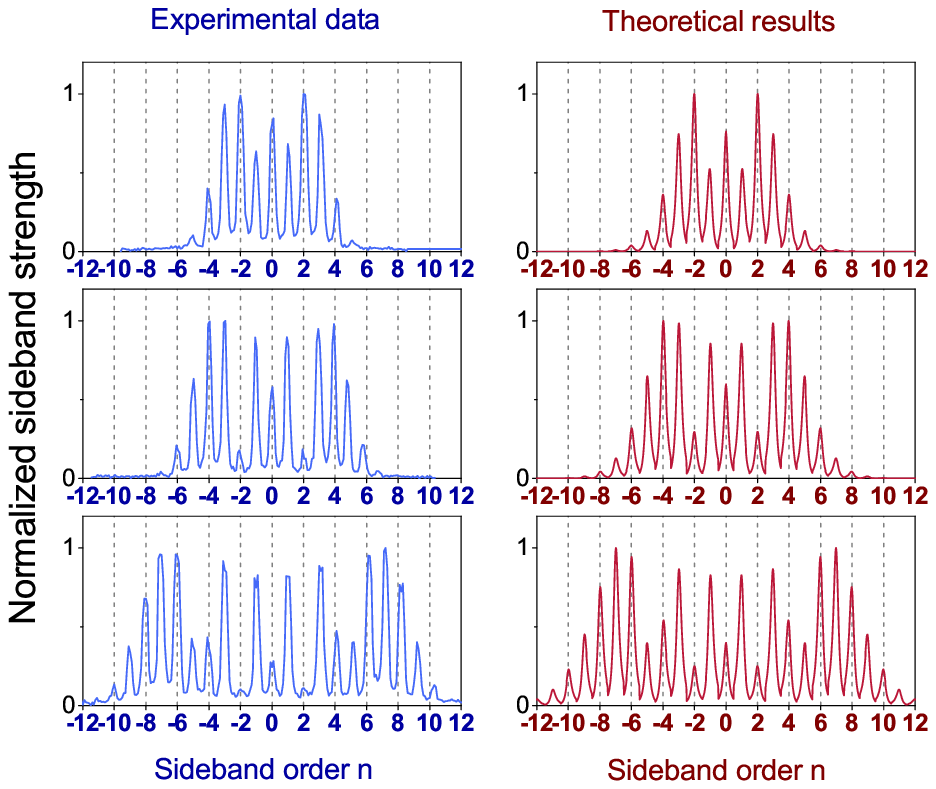}
\caption{Quantitative comparisons of high-order sideband spectra for different modulation amplitudes between the experimental data and the theoretical results. For the experimental data (left column), the parameter $\omega_a$ is taken to be $0.42\times2\pi$, $0.64\times2\pi$, and $1.08\times2\pi$ GHz from top to bottom, respectively. While for the theoretical plots (right column), the corresponding fitting parameter $\alpha$ takes the values of 3.3, 4.85, and 8.23, respectively. The shaking rate $\omega_s$ is set to be $50\times2\pi$ Hz. In the theoretical calculations, a decay rate of $\gamma_2$ for the two-level transition is used.}
\label{figure3}
\end{figure}

Figure \ref{figure3} depicts the dependence of the strength distribution of the high-order sideband peaks on the modulation amplitude $\omega_a$ with a fixed shaking rate $\omega_s$ for the optical lattice. For each measured high-order sideband spectrum on the left column in Fig. \ref{figure3} with a given $\omega_a$ value, a fitting value of $\alpha$ can be identified to reproduce the experimental data by plotting the theoretical distribution function given in Eq. (\ref{sideband}), as shown in the right column. A careful comparison of the strength distributions in the high-order sideband spectra shows that the simple theoretical model with an analytical solution, as given in Eq. (\ref{sideband}), can well explain the experimentally observed data, even to a fine-detail level.

\begin{figure}[ptb]
\centering
\includegraphics[width=8.5cm]{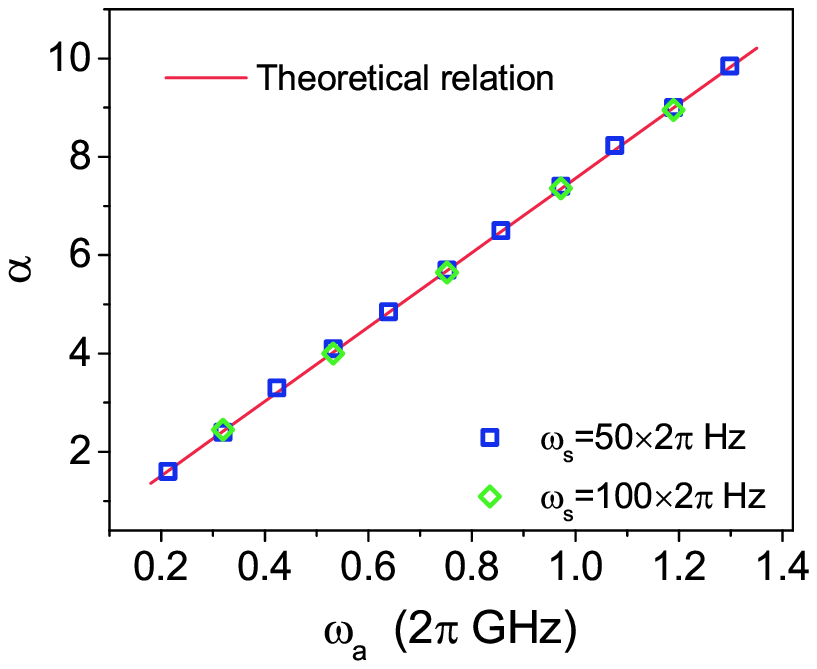}
\caption{Comparison of the parameter $\alpha$ between the theoretical expression and the fitting values from the experimentally measured spectra. The theoretical relation between $\alpha$ and $\omega_a$ is drawn in (red) line, in which the length $L$ is taken to be 0.31 m. The fitting values of $\alpha$ are obtained by requiring the best matches of the theoretical and experimental spectra, as shown in Fig. \ref{figure3}. The (blue) squares represent the values from the spectra measured when $\omega_s$ is $50\times2\pi$ Hz, and (green) diamonds are the ones when $\omega_s$ is $100\times2\pi$ Hz.}
\label{figure4}
\end{figure}

To better test the theoretical model quantitatively, a careful comparison is performed between the theoretically determined parameter results of $\alpha$ and the experimentally fitted values at different scanning rates. Since $\alpha=\frac{\omega_a\omega_0L}{\omega_Lc}$ is given by the theory and all other parameters can be determined independently, the linear dependence of $\alpha$ on $\omega_a$ can be established, as drawn by the red line in Fig. \ref{figure4}. The fitting values by best reproducing the experimental spectra, as represented by examples shown in Fig. \ref{figure3}, are also marked in Fig. \ref{figure4}, including two sets of results under different shaking rates. Such perfect consistency clearly indicates that our experimental setup is indeed an ideal model system with perfect two-level atoms inside a shaken optical lattice. This system works in the unique parametric regime of having the shaking rate to be far smaller than the energy gap of the harmonic trapping potential but still much larger than the linewidth of the two-level system, which is ideal to study various quantum dynamical behaviors, such as the high-order sideband effect shown above.

Similar high-order sideband generations (HSG) in semiconductors were reported recently in the semiconductor quantum wells \cite{Zaks2012}, bulk GaAs \cite{Zaks2013}, and tungsten diselenide \cite{Langer2016}. The physical mechanism underlying the HSG observed in semiconductors is quite different from that governing our current two-level atomic system. In those semiconductor systems, the excitons are first excited, which then absorb energy as they are accelerated by an external terahertz field and finally generate the high-order sidebands of the original excitation.\cite{Yan2008, Yan2017} Because of the inevitable scattering, the excitonic decay rate approaches an order close to the terahertz level, and the frequency of the periodic field must be larger than this decay rate to clearly observe the HSG effect. Theories in \cite{Yan2008, Yan2017, Crosse2014} show that the terahertz field strength must reach the order of $1-10$ kV/cm, which means a free electron laser has to be employed. However, the high-order sideband effect can be easily observed in our system primarily because the linewidth achieved in the dipole-forbidden transition of $^{87}$Sr atom is so narrow that it permits the shaking rate to be on the order of 100 Hz while still satisfying the condition of $\omega_s\gg\gamma_2$. In such case, the effective ac field strength required for the observation of high-order sidebands is determined by the modulation amplitude, which is easily accessible in the experiment with a low power diode laser. Meanwhile the relatively low shaking rate of the lattice field does not disturb the energy structure of the harmonic trapping potential (with $\omega_s\ll\omega_p$). Such shaken optical lattice introduces a Doppler effect relative to the excitation laser beam for the two-level atomic transition, which provides the required ac field to generate the high-order sideband effect. This is quite different from the mechanism of accelerating excitons in an external field for semiconductors. Furthermore, since two-level atoms do not have the energy dispersion as the excitons do in semiconductors, our current system exhibits new features in the sideband spectrum.

In general, the observed high-order sideband spectra in the system of two-level atoms inside a shaken optical lattice have the following distinct characteristics: (1) The HSE features a series of discrete and equidistant sideband peaks with the same frequency interval of $\omega_s$, which follows the requirement of energy conservation for the involved photons. (2) Both even- and odd-order sideband peaks occur in each spectrum, which is a unique feature of the high-order sideband effect in the dispersion-less system, while only even-order sideband peaks observed in semiconductors because of the symmetry in the momentum space (contributions from momenta $k$ and $-k$ would cancel each other out\cite{Yan2008, Yan2017} for odd-order sidebands). (3) In the current atomic system, the high-order sideband spectrum presents a symmetrical order, i.e., the $n$-th and -$n$-th order sidebands have the same intensity weight. In semiconductors, the strengths of the negative-order sidebands decay more rapidly than the positive-order sidebands as the sideband number increases; therefore, only a few sideband peaks can be observed.

Using the selected dipole-forbidden transition in $^{87}$Sr atoms loaded inside a shaken optical lattice, we have realized an ideal model system with perfect two-level atoms driven by a time-dependent periodic field, and observed the interesting high-order sideband effect which began to draw attentions in semiconductors recently. The shaken optical lattice introduces an effective ac field to act on the two-level atoms through the Doppler effect, but still keeps the energy structure of the lattice potential unchanged. High-order sideband peaks up to at least ¡À20 orders have been experimentally observed when the modulation amplitude for the lattice field increases to a certain value. Several new features appear in this new model system with two-level atoms in the shaken optical lattice as compared to the HSG seen in semiconductors. The experimentally observed phenomena can be well explained by the exact analytical solution of our theory. The use of the dipole-forbidden transition with an ultra-narrow linewidth to form an ideal closed two-level system allows us to access new operating regimes in the arena of atoms in shaken optical lattices and they can be used to simulate dynamics of condensed-matter physics systems and lead to exciting new discoveries. Actually, by selecting different two-level transitions in the atoms (varying the atomic decay rate $\gamma_2$) and tuning the shaking rate ($\omega_s$), as well as the modulation amplitude ($\omega_a$), one can investigate broad parametric regions for this model system and explore many interesting quantum dynamical effects. The observed high-order sidebands can expand the single-frequency quantum bit, as proposed for a two-level system inside a shaken optical lattice, into a multiplexer \cite{Schneider2012}, which would significantly increase its potential applications in quantum computation.

This work was supported by the National Natural Science Foundation of China (Grant Nos. 11474282 and 61435007), the Key Research Project of Frontier Science of the Chinese Academy of Sciences (Grant No. QYZDB-SSW-JSC004) and the Strategic Priority Research Program of the Chinese Academy of Sciences (Grant No. XDB21030100).

Additional information:See supplementary information.\\
$\dag$These authors contributed equally to this work.\\
*Corresponding authors with emails: jyyan@bupt.edu.cn, mxiao@uark.edu, changhong@ntsc.ac.cn.


\begin{thebibliography}{99}
\bibitem{Allen1975} L. Allen and J. H. Eberly, Optical Resonance and Two-Level Atoms (John Wiley, 1975).

\bibitem{Eckardt2017} A. Eckardt, Rev. Mod. Phys. {\bf 89}, 011004 (2017).

\bibitem{Greiner2008} M. Greiner and S. F\"{o}lling, Nature {\bf 453}, 736-738 (2008).

\bibitem{Bloch2008} I. Bloch, J. Dalibard and W. Zwerger, Rev. Mod. Phys. {\bf 80}, 885-964 (2008).

\bibitem{Goldman2016} N. Goldman, J. C. Budich and P. Zoller, Nat. Phys. {\bf 12}, 639¨C645 (2016).

\bibitem{Gadway2012} B. Gadway, D. Pertot, J. Reeves and D. Schneble, Nat. Phys. {\bf 8}, 544-549 (2012).

\bibitem{Kolkowitz2017} S. Kolkowitz, S. L. Bromley, T. Bothwell, M. L. Wall, G. E. Marti, A. P. Koller, X. Zhang, A. M. Rey and J. Ye, Nature {\bf 542}, 66-70 (2017).

\bibitem{Lignier2007} H. Lignier, C. Sias, D. Ciampini, Y. Singh, A. Zenesini, O. Morsch and E. Arimondo, Phys. Rev. Lett. {\bf 99}, 220403 (2007).

\bibitem{Lundblad2008} N. Lundblad, P. J. Lee, I. B. Spielman, B. L. Brown, W. D. Phillips and J. V. Porto, Phys. Rev. Lett. {\bf 100}, 150401 (2008).

\bibitem{Tokuno2011} A. Tokuno and T. Giamarchi, Phys. Rev. Lett. {\bf 106}, 205301 (2011).

\bibitem{Struck2012} J. Struck, C. \"{O}lschl\"{a}ger, M. Weinberg, P. Hauke, J. Simonet, A. Eckardt, M. Lewenstein, K. Sengstock and P. Windpassinger, Phys. Rev. Lett. {\bf 108}, 225304 (2012).

\bibitem{Parker2013} C. V. Parker, L.-C. Ha and C. Chin, Nat. Phys. {\bf 9}, 769-774 (2013).

\bibitem{Ha2015} L.-C. Ha, L. W. Clark, C. V. Parker, B. M. Anderson and C. Chin, Phys. Rev. Lett. {\bf 114}, 055301 (2015).

\bibitem{Clark2016} L. W. Clark, L. Feng and C. Chin, Science {\bf 354}, 606-610 (2016).

\bibitem{Gutterres2002} R. F. Gutterres, C. Amiot, A. Fioretti, C. Gabbanini, M. Mazzoni and O. Dulieu, Phys. Rev. A {\bf 66}, 024502 (2002).

\bibitem{Sell2011} J. F. Sell, B. M. Patterson, T. Ehrenreich, G. Brooke, J. Scoville and R. J. Knize, Phys. Rev. A {\bf 84}, 010501 (2011).

\bibitem{Juska2013} G. Juska, V. Dimastrodonato, L. O. Mereni, A. Gocalinska and E. Pelucchi, Nat. Photon. {\bf 7}, 527-531 (2013).

\bibitem{Keil2017} R. Keil, M. Zopf, Y. Chen, B. H\"{o}fer, J. Zhang, F. Ding and O. G. Schmidt, {\bf 8}, 15501 (2017).

\bibitem{Liu2007} R.-B. Liu and B.-F. Zhu, AIP Conference Pro. {\bf 893}, 1455 (2007).

\bibitem{Yan2008} J.-Y. Yan, Phys. Rev. B {\bf 78}, 075204 (2008).

\bibitem{Zaks2012} B. Zaks, R. B. Liu and M. S. Sherwin, Nature {\bf 483}, 580-583 (2012).

\bibitem{Zaks2013} B. Zaks, H. Banks and M. S. Sherwin, Appl. Phys. Lett. {\bf 102}, 012104 (2013).

\bibitem{Langer2016} F. Langer, M. Hohenleutner, C. P. Schmid, C. Poellmann, P. Nagler, T. Korn, C. Sch\"{u}ller, M. S. Sherwin, U. Huttner, J. T. Steiner, S. W. Koch, M. Kira and R. Huber, Nature {\bf 533}, 225-229 (2016).

\bibitem{Schneider2012} P.-I. Schneider and A. Saenz, Phys. Rev. A {\bf 85}, 050304 (2012).

\bibitem{Takamoto2003} M. Takamoto and H. Katori, Phys. Rev. Lett. {\bf 91}, 223001 (2003).

\bibitem{Campbell2017} S. L. Campbell, R. B. Hutson, G. E. Marti, A. Goban, O. N. Darkwah, R. L. McNally, L. Sonderhouse, J. M. Robinson, W. Zhang, B. J. Bloom and J. Ye, Science {\bf 358}, 90-94 (2017).

\bibitem{Bergquist1986} J. C. Bergquist, R. G. Hulet, W. M. Itano and D. J. Wineland, Phys. Rev. Lett. {\bf 57}, 1699-1702 (1986).

\bibitem{Yan2017} J.-Y. Yan, J. Appl. Phys. {\bf 122}, 084306 (2017).

\bibitem{Crosse2014} J. A. Crosse, X. Xu, M. S. Sherwin and R. B. Liu, Nat. Commun. {\bf 5}, 4854-4859 (2014).


\end{thebibliography}
\end{document}